\documentstyle[epsf]{cupconf}

\def\etal{{\it et al.}}
\font\ninerm=cmr9
\def\cptn #1{{\noindent{\baselineskip 8pt \ninerm #1}}}

\def\eps@scaling{.95}
\def\epsscale#1{\gdef\eps@scaling{#1}}

\def\plotone#1{\centering \leavevmode
    \epsfxsize=\eps@scaling\columnwidth \epsfbox{#1}}

\begin{document}
\input psfig.tex

\title[Globular-Cluster CMDs]{Globular-Cluster
Color--Magnitude Diagrams with HST\footnote{Based on observations with the
NASA/ESA {\it Hubble Space Telescope}, obtained at the Space Telescope
Science Institute, which is operated by AURA, Inc., under NASA contract
NAS5-26555}}

\author[C.\ Sosin {\it et al.}]{%
 C.\ns S\ls O\ls S\ls I\ls N$^1$,\ns
 G.\ns P\ls I\ls O\ls T\ls T\ls O$^2$,\ns
 S.\ns G.\ns D\ls J\ls O\ls R\ls G\ls O\ls V\ls S\ls K\ls I$^3$,\\
 I.\ns R.\ns K\ls I\ls N\ls G$^1$,\ns
 R.\ns M.\ns R\ls I\ls C\ls H$^4$,\ns
 B.\ns D\ls O\ls R\ls M\ls A\ls N$^5$,\\
 J.\ns L\ls I\ls E\ls B\ls E\ls R\ls T$^6$,\ns
 \and\ A.\ns R\ls E\ls N\ls Z\ls I\ls N\ls I$^7$}

\affiliation{$^1$Astronomy Dept., University of California, Berkeley, CA
 92720-3411, USA\\[\affilskip]
 $^2$Dipartimento di Astronomia, Universit\`a di Padova, Vicolo
 dell' Osservatorio 5, I-35122 Padova, Italy\\[\affilskip]
 $^3$Astronomy Dept., MS 105-24, California Institute of 
 Technology, Pasadena, CA 91125, USA\\[\affilskip]
 $^4$Astronomy Dept., Columbia University, 538 W.\ 120th St., Box 43
 Pupin, New York, NY 10027, USA\\[\affilskip]
 $^5$Laboratory for Astronomy and Solar Physics, Code 681, NASA Goddard
 Space Flight Center, Greenbelt, MD 20771, USA\\[\affilskip]
 $^6$Steward Observatory, University of Arizona, Tucson, AZ 85721, 
 USA\\[\affilskip]
 $^7$Dipartimento di Astronomia, Universit\`a di Bologna, Cp.\ 596, 
 I-40100 Bologna, Italy}
\makepptitle

\begin{abstract}
This poster paper illustrates the color--magnitude diagrams discussed by
Piotto \etal\ in the preceding paper.  We present CMDs for 13 clusters;
and we emphasize the discovery of additional blue horizontal branches in
two metal-rich clusters, and the four-mode HB of NGC 2808.
\end{abstract}

\firstsection
\section{Introduction} 

The factors that determine the morphology of the horizontal branch in
globular clusters are still not well understood.  Metal abundance plays
an important role, with the most metal-rich clusters usually having the
reddest HBs.  However, a number of ``second parameters,'' such as age,
have been proposed as explanations of the clusters that deviate from
this rule.

Here we present preliminary results from a {\it Hubble Space Telescope}
program that aims to explore connections between stellar evolution and
cluster dynamics, and to investigate the CMD morphology of some clusters
that are difficult to observe from the ground.  The central regions of
ten clusters were observed with the WFPC2 on {\sl HST}.  Some of the
clusters chosen had a high central density and/or concentration; others
had ultraviolet flux of unknown origin detected by {\sl IUE}.  We also
include here three additional CMDs that we have derived from archival
data originally taken by Yanny in another program.

We present two results:\ the discovery of additional blue horizontal
branches in two metal-rich globular clusters, and the 
intriguing ``clumpy'' nature of the blue horizontal-branch tail of NGC
2808.

\section{Blue horizontal branches in metal-rich clusters}

An exciting result from this program has been the discovery of
additional {\it blue} horizontal branches in two {\it metal-rich}
clusters, NGC 6388 and NGC 6441.  Both are extremely crowded from the
ground, and their HBs have not been seen previously.

Both clusters were also detected in the ultraviolet by {\sl IUE}, and it
was suggested that their UV brightness arose from blue HB stars.  While
a handful of blue HB stars have indeed been detected in metal-rich
populations, blue populations as large as those discovered here have not
previously been seen.  (For discussion of these points, and references,
see the preceding paper by Piotto \etal.)  What is puzzling, however, is
that some other clusters with nearly the same metal abundance have no
blue HB stars (see our CMDs of 47 Tuc and NGC 5927).  It appears that
another ``second-parameter'' problem may be at hand!

\hbox{
\vbox{\hsize 2.5 truein
\psfig{figure=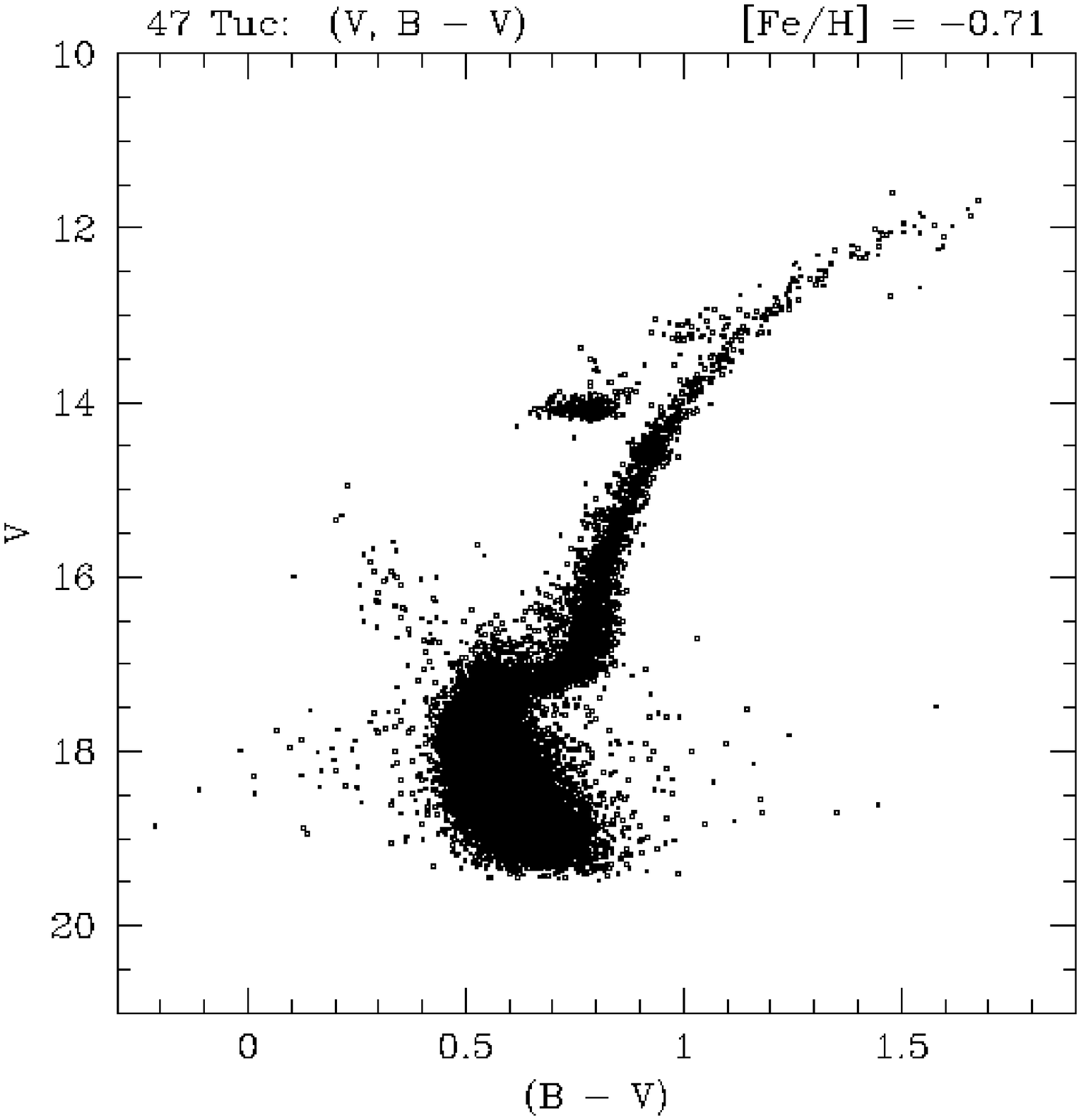,height=2.6truein,width=2.6truein}
}
\vbox{\hsize 2.5 truein
\psfig{figure=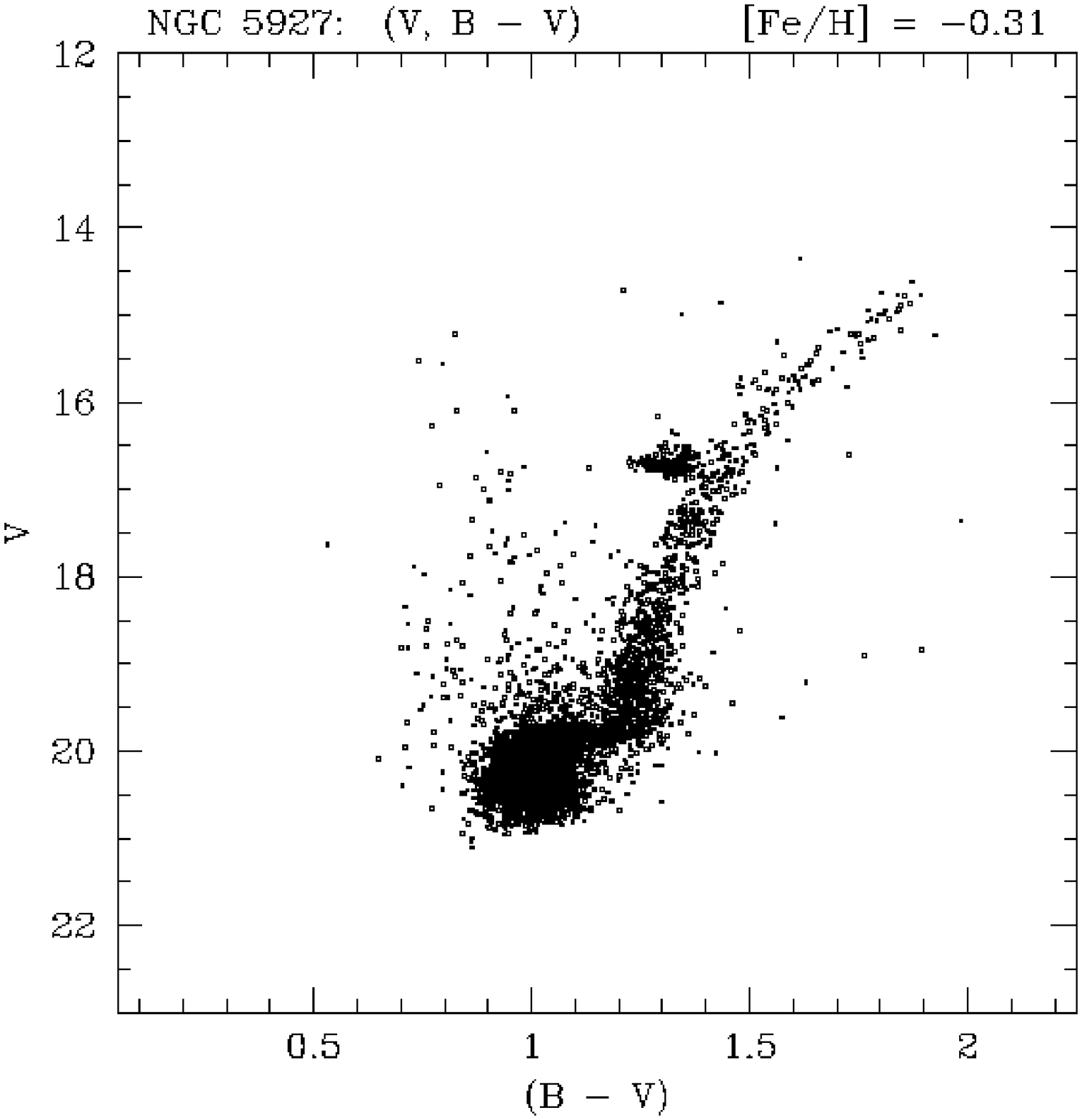,height=2.6truein,width=2.6truein}
}
}

\hbox{
\vbox{\hsize 2.5 truein
\psfig{figure=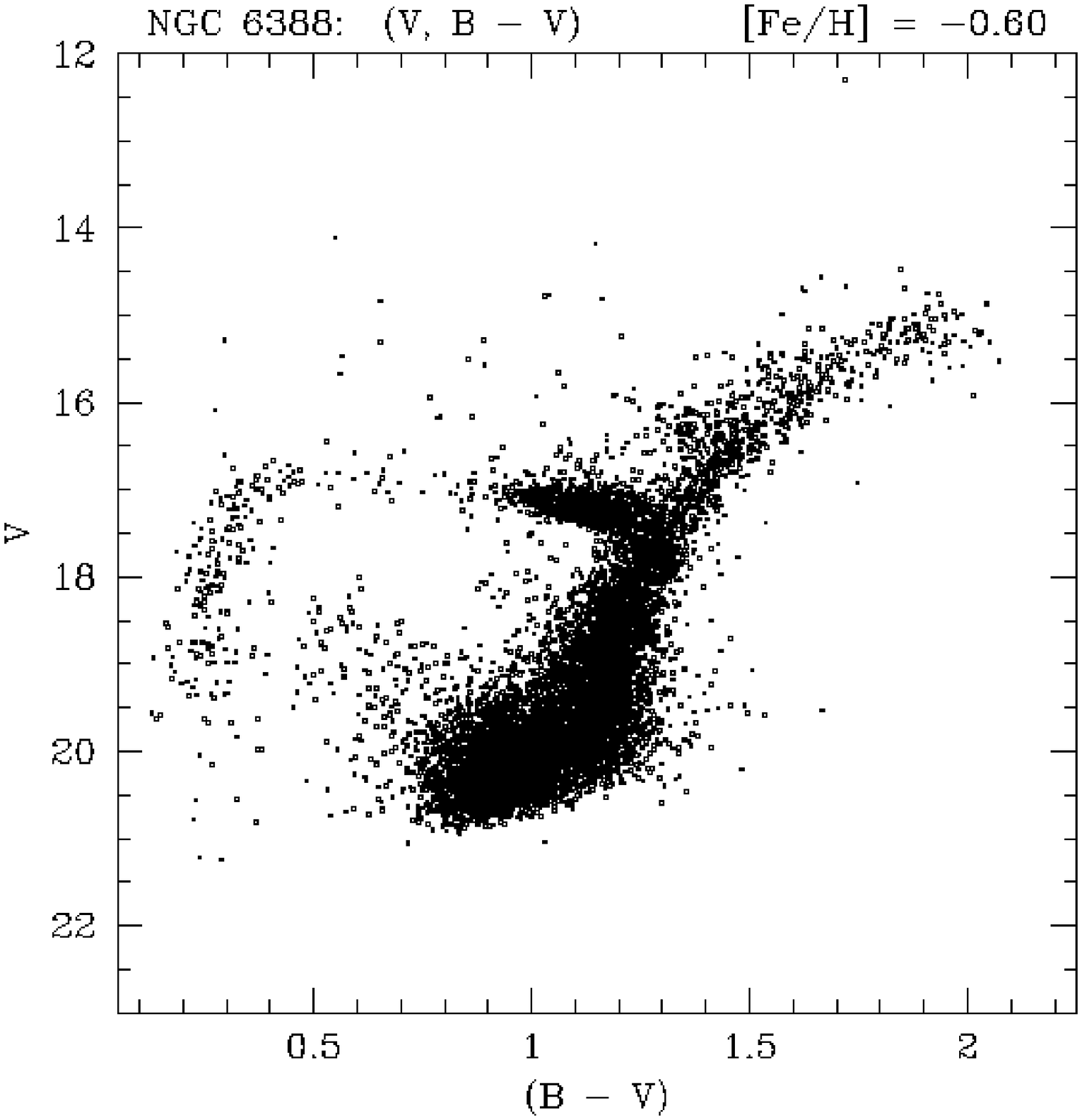,height=2.6truein,width=2.6truein}
}
\vbox{\hsize 2.5 truein
\psfig{figure=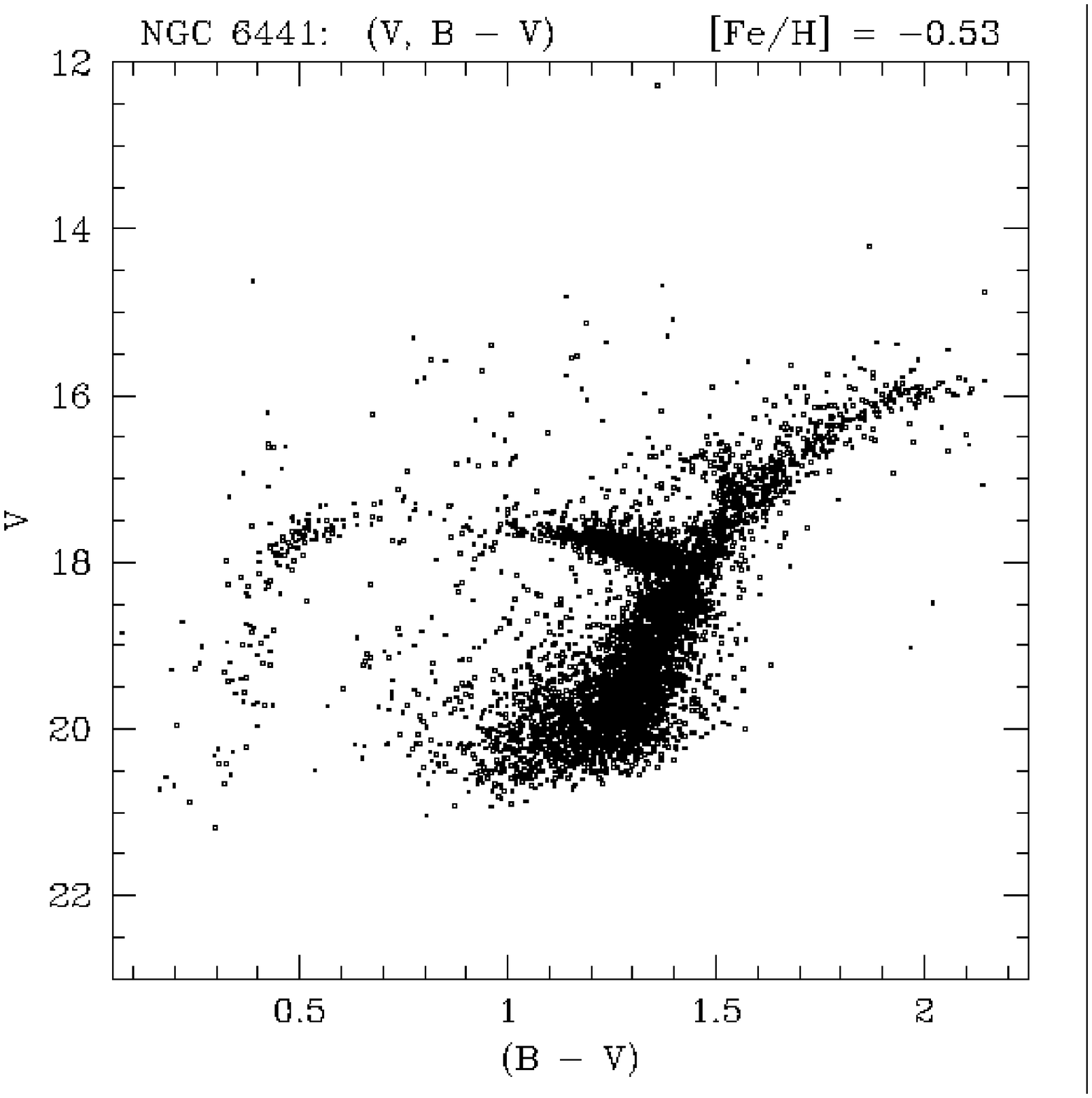,height=2.6truein,width=2.6truein}
}
}
\cptn{Figure 1. The upper two high-metal-abundance clusters have only a red
stub of a horizontal branch, while the lower two have an additional blue
portion.}

\section{The Long, Clumpy HB of NGC 2808}

The horizontal branch of NGC 2808 has provided another fascinating
result.  Already known to be separated from the red stub, its blue end
is now seen to extend many magnitudes fainter in $V$ than previously
known, down to $V \simeq 21$.

A CMD taken with an ultraviolet filter (F218W, $\lambda_{\rm eff} =
2189 {\rm \AA}$) spreads out the HB tail in color, allowing better
separation of stars of different envelope mass.  A histogram of this
``blue vertical branch'' shows that it is made up of at least two
distinct groups, and possibly a third at its extreme blue end (in
addition to the red stub that does not appear in the UV diagram).  No
mass-loss mechanism is known that could lead to such well-defined HB
clumps.

\hbox{
\vbox{\hsize 2.5 truein
\psfig{figure=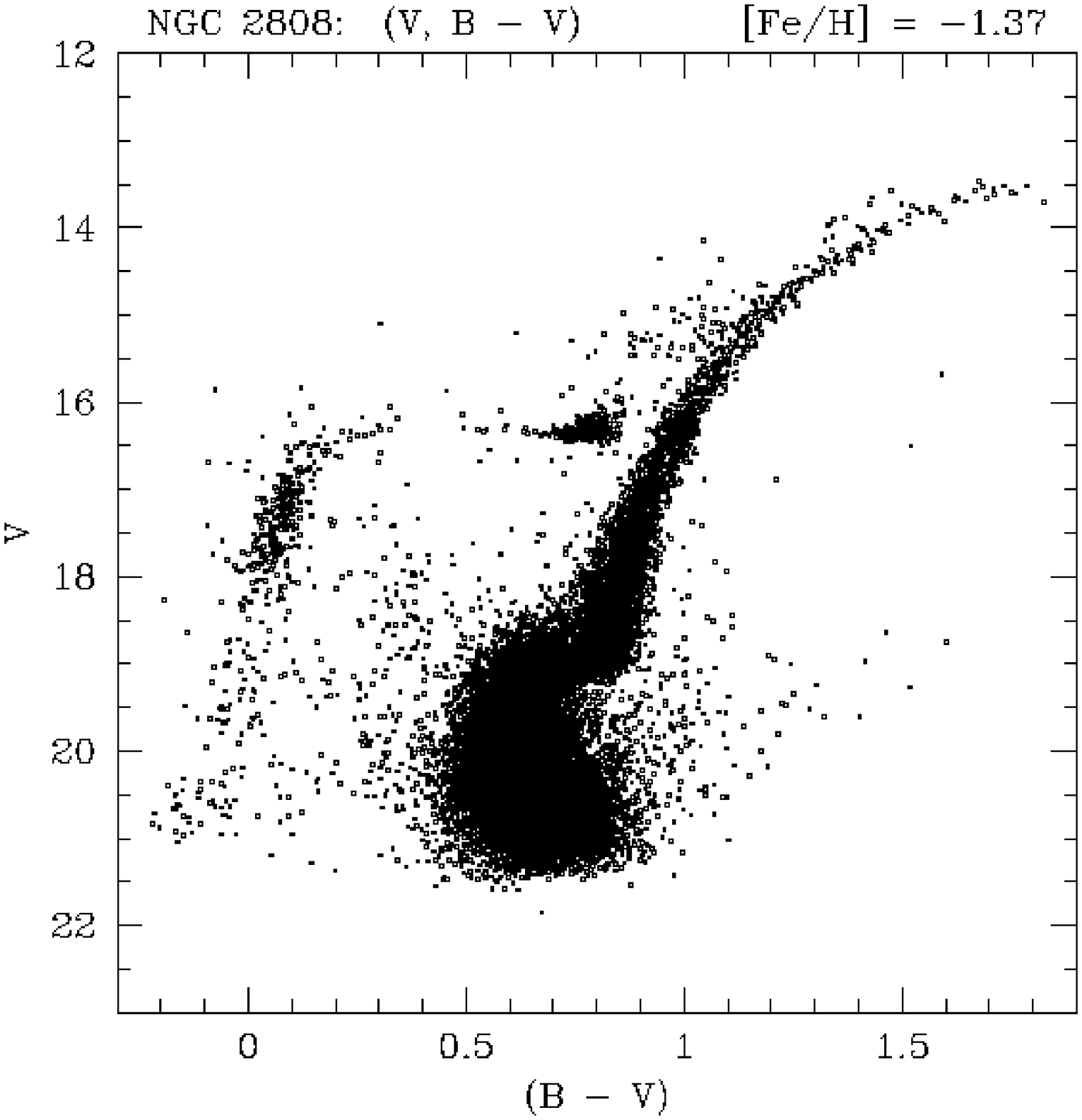,height=2.6truein,width=2.6truein}
}
\vbox{\hsize 2.5 truein
\psfig{figure=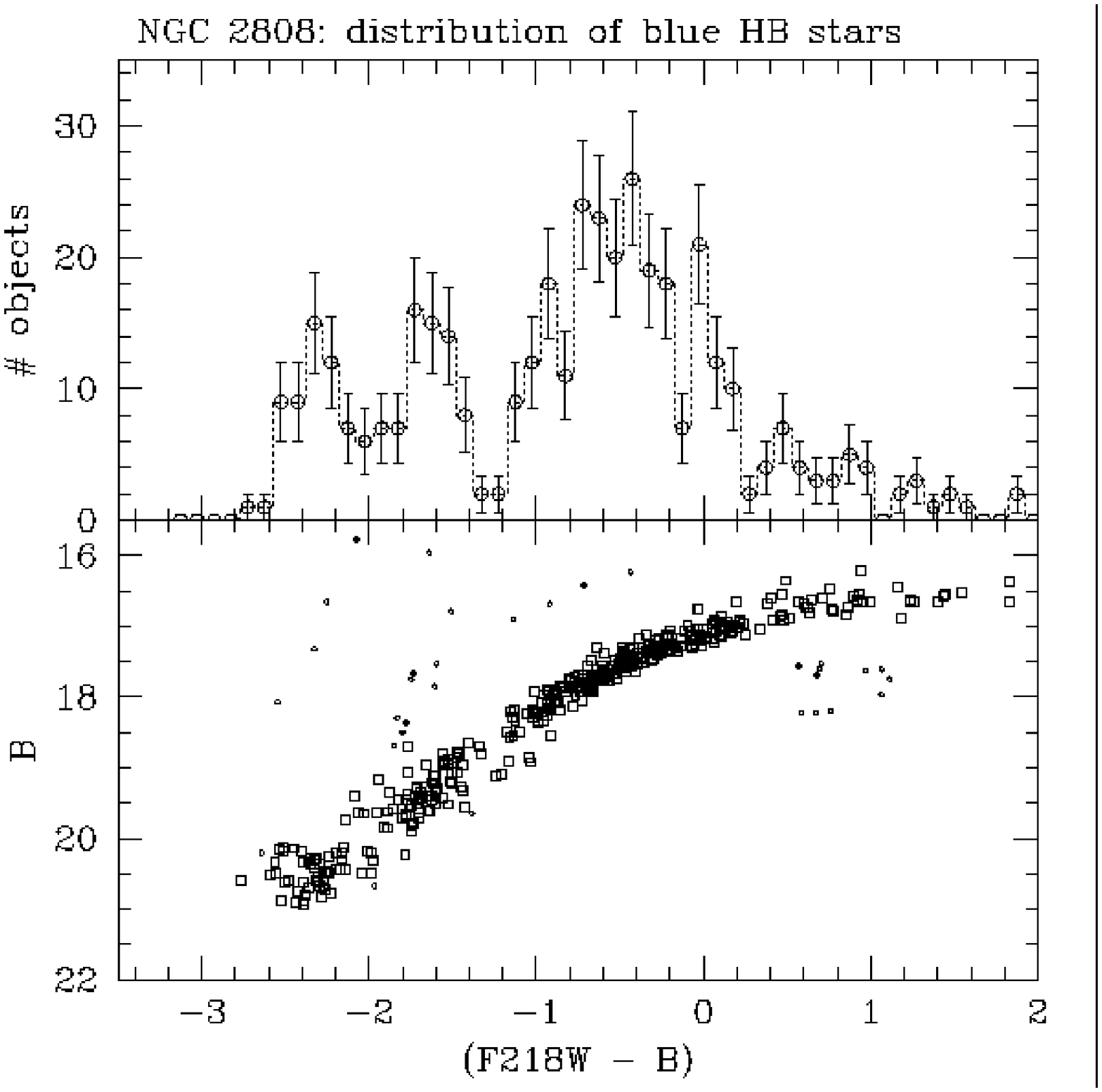,height=2.6truein,width=2.6truein}
}
}
\cptn{Figure 2. (left) The ($B$, $V$) CMD of NGC 2808.  (right) The
(FUV, $B$) CMD of NGC 2808, with a graph above it showing the
distribution of the stars in color.  Note that the red-stub part of the 
HB is too weak in the FUV to appear in this diagram at all.}

\section{Other clusters}

The analysis of these data continues.  Color--magnitude diagrams for
our other clusters are shown in Figure 3; a number of
interesting features in them are being investigated.

\hbox{
\vbox{\hsize 2.5 truein
\psfig{figure=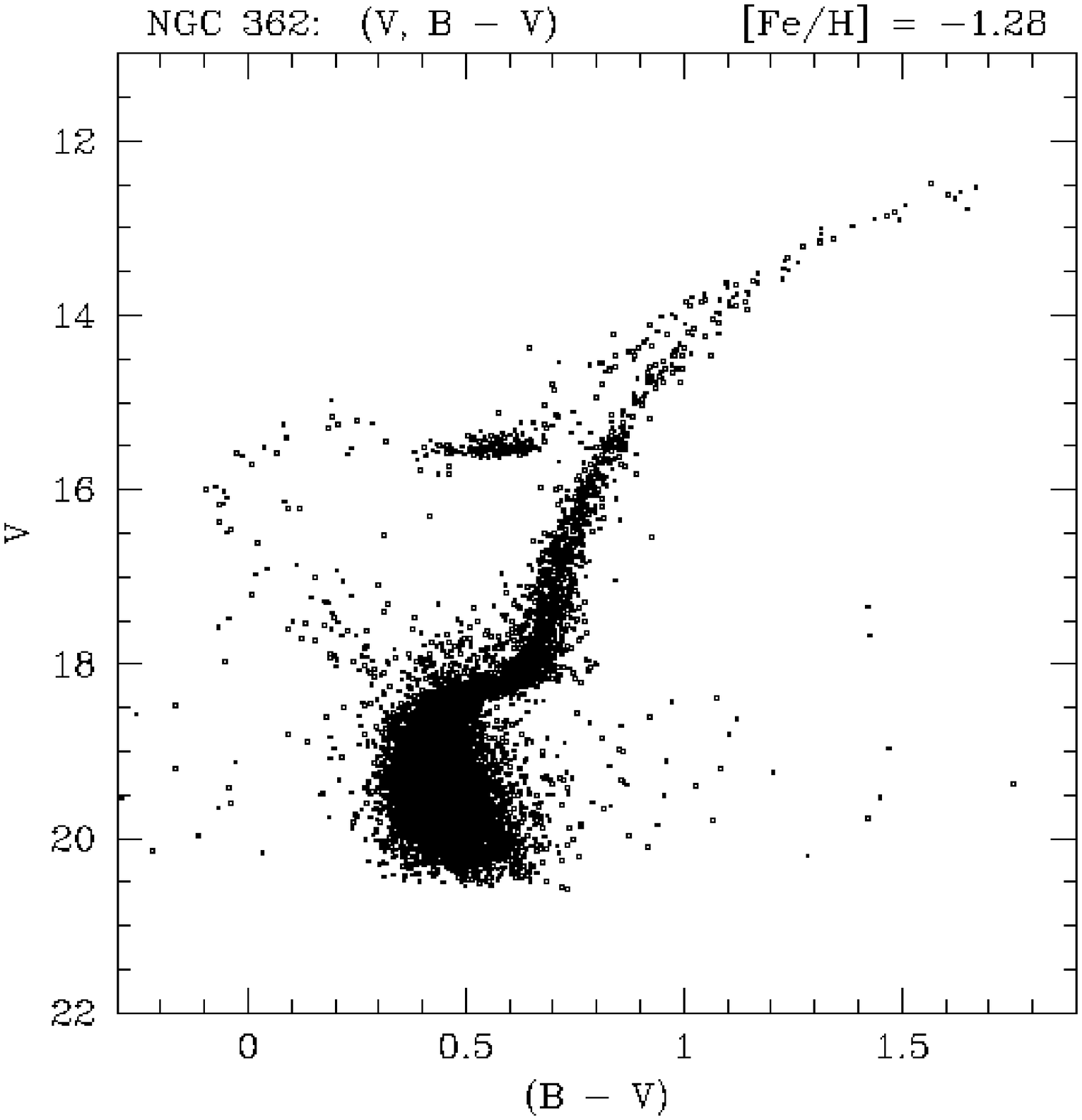,height=2.6truein,width=2.6truein}
}
\vbox{\hsize 2.5 truein
\psfig{figure=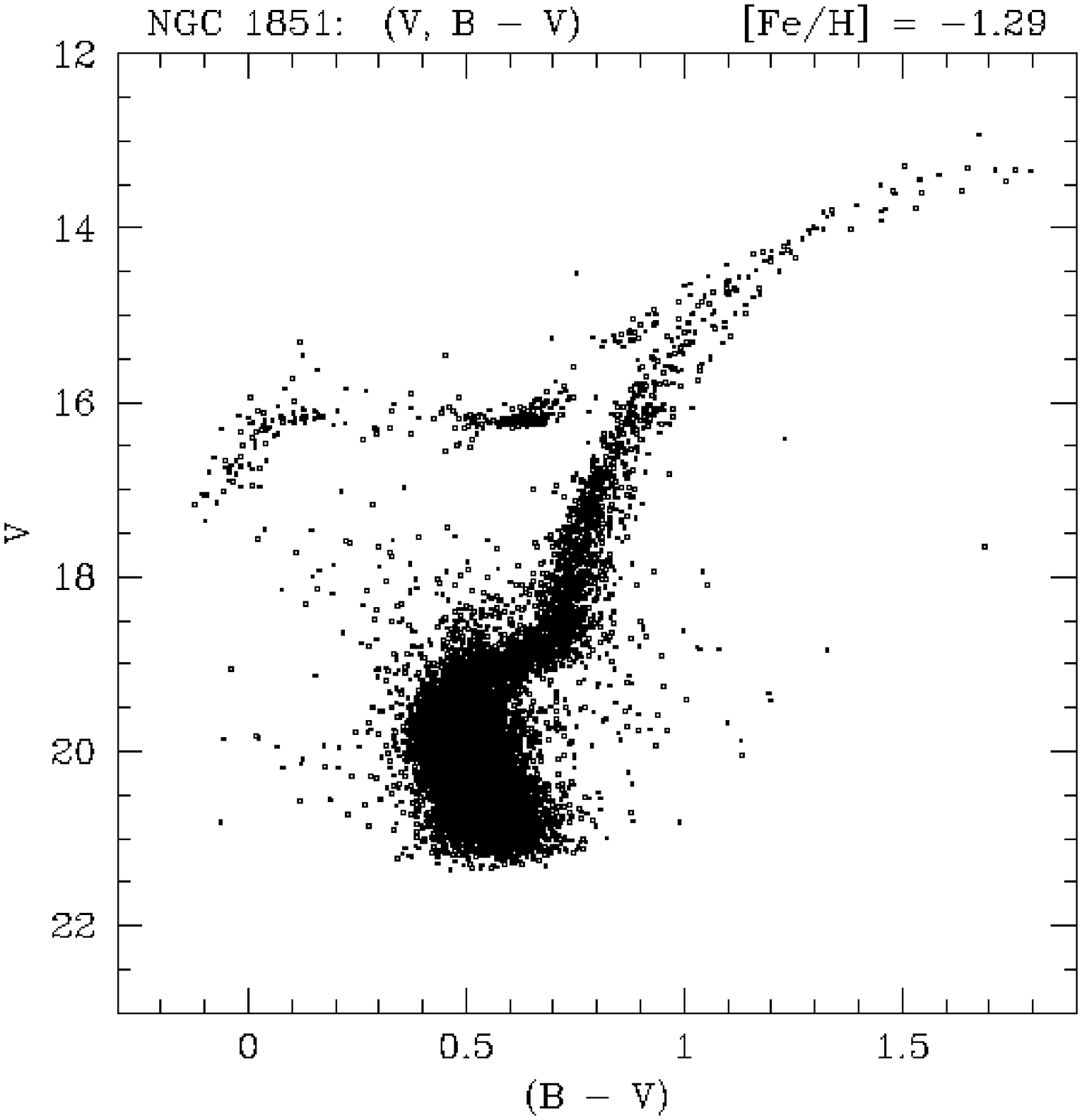,height=2.6truein,width=2.6truein}
}
}

\hbox{
\vbox{\hsize 2.5 truein
\psfig{figure=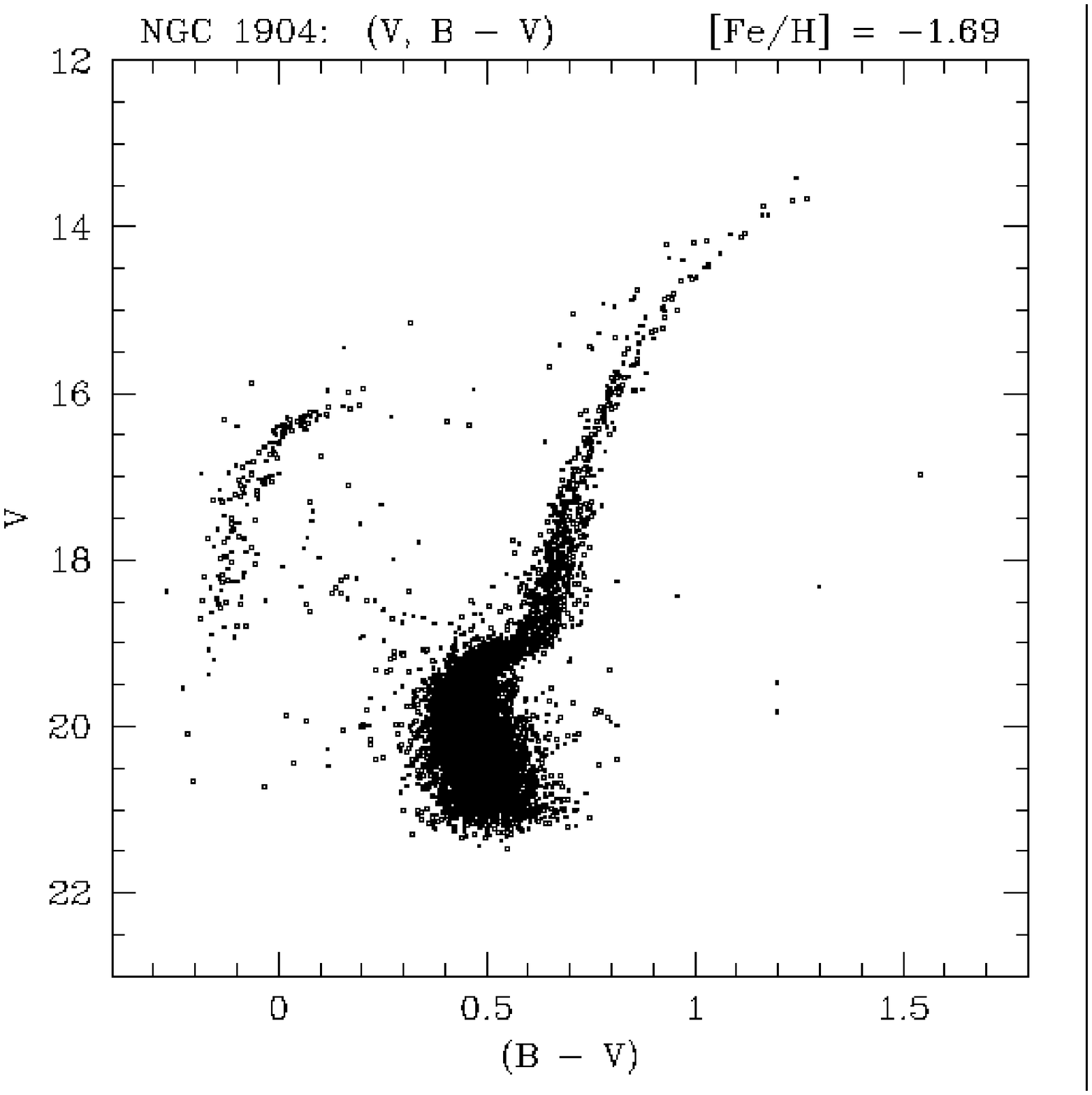,height=2.6truein,width=2.6truein}
}
\vbox{\hsize 2.5 truein
\psfig{figure=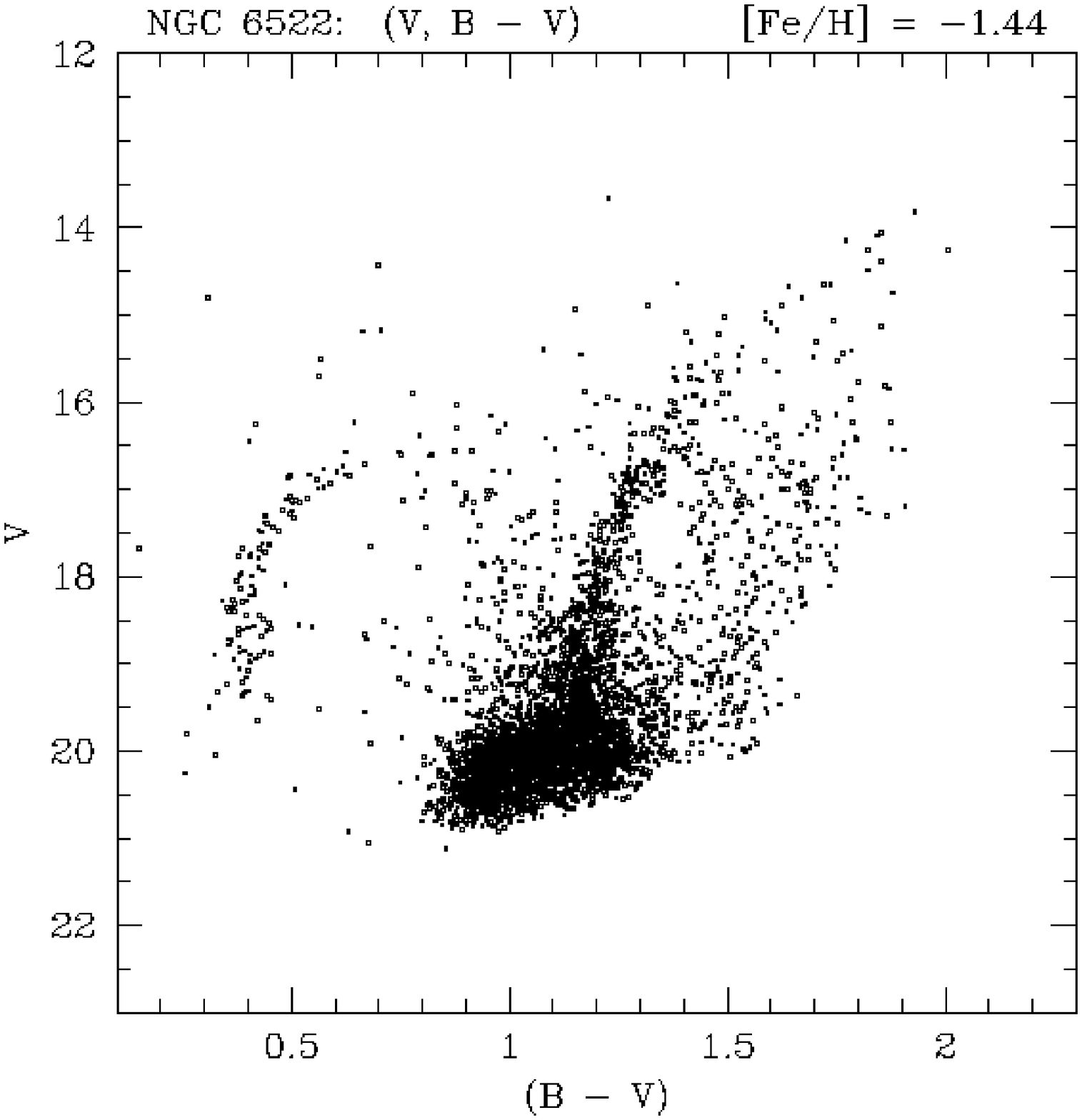,height=2.6truein,width=2.6truein}
}
}

\hbox{
\vbox{\hsize 2.5 truein
\psfig{figure=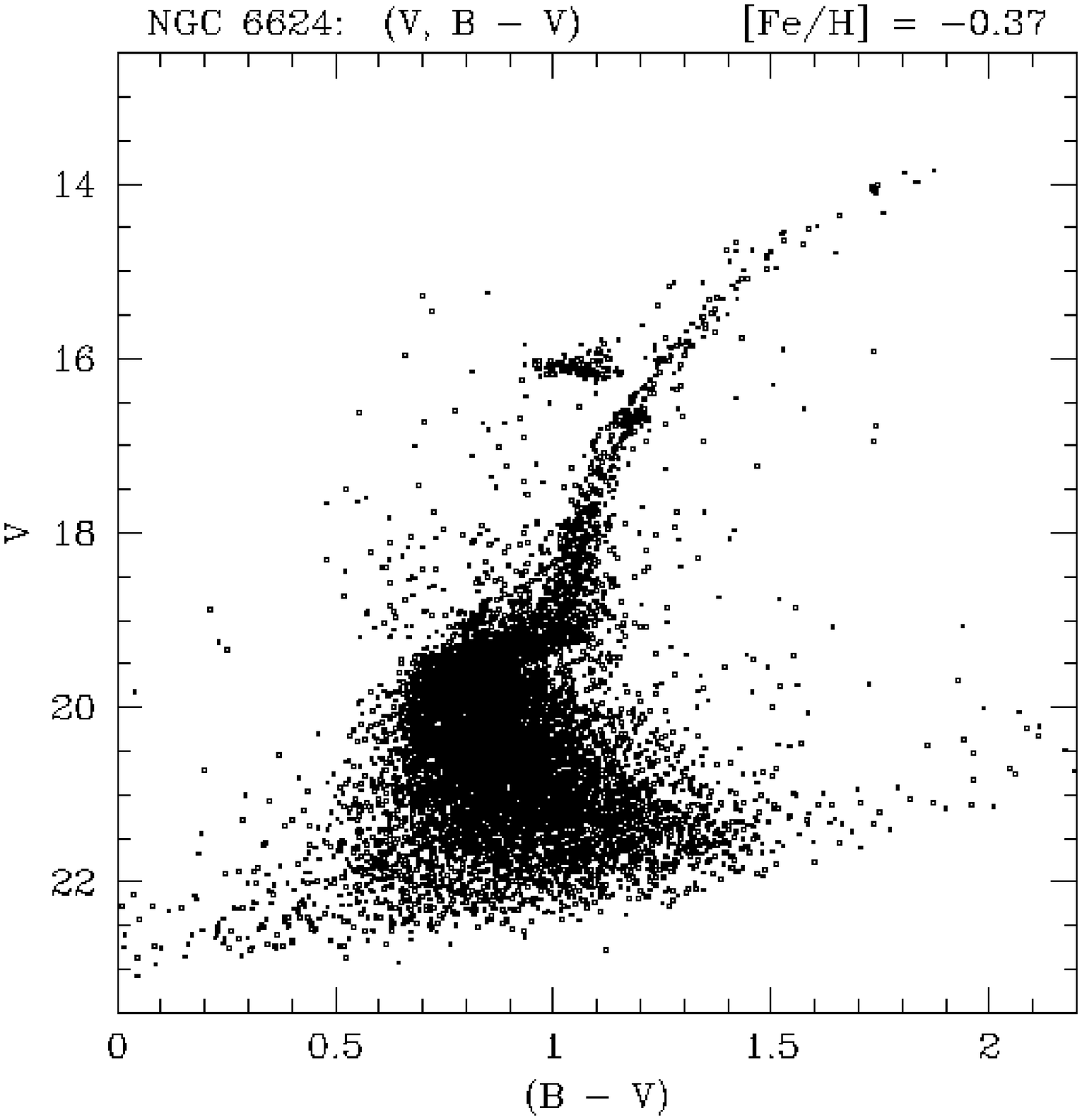,height=2.6truein,width=2.6truein}
}
\vbox{\hsize 2.5 truein
\psfig{figure=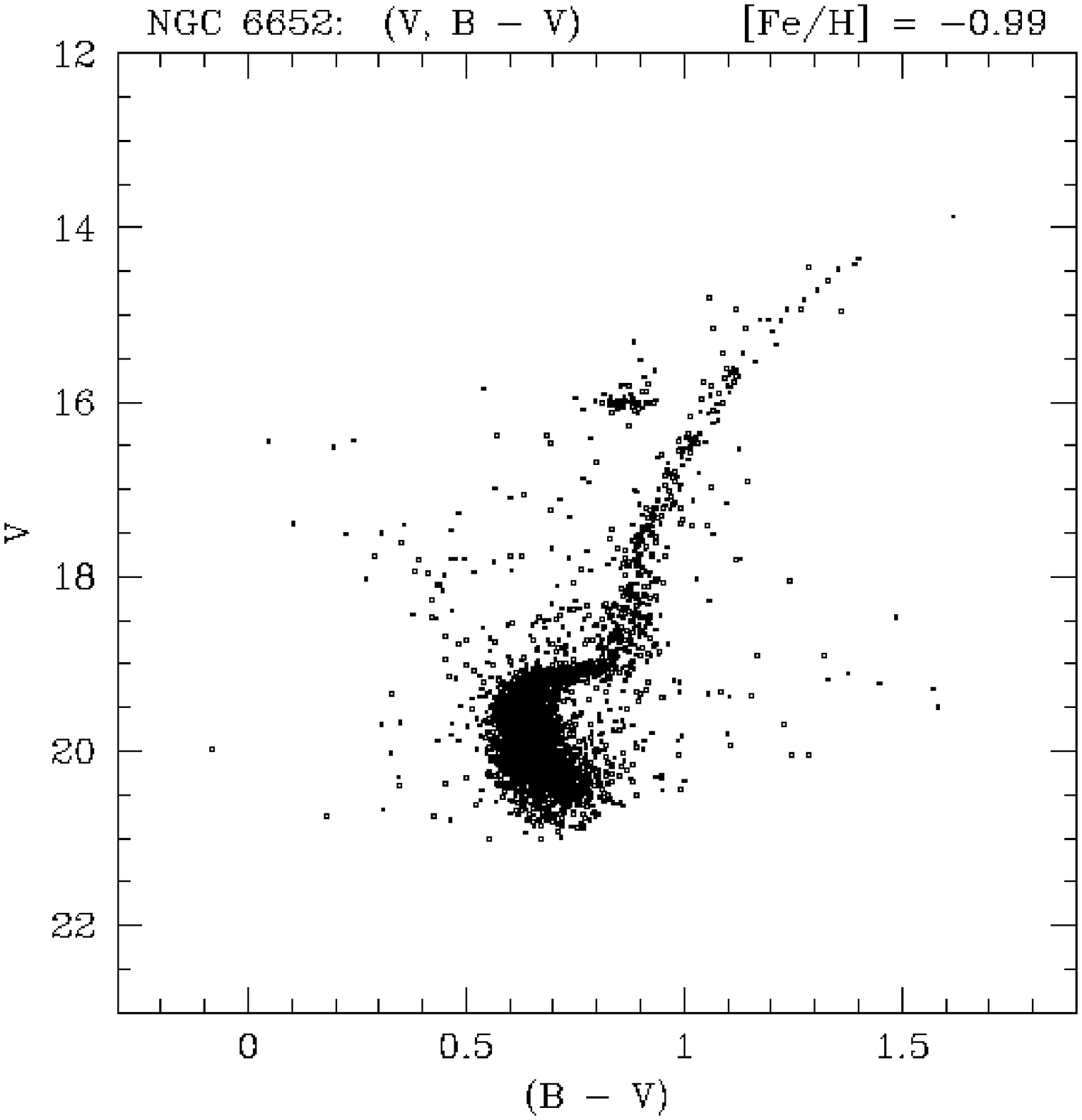,height=2.6truein,width=2.6truein}
}
}

\hbox{
\vbox{\hsize 2.5 truein
\psfig{figure=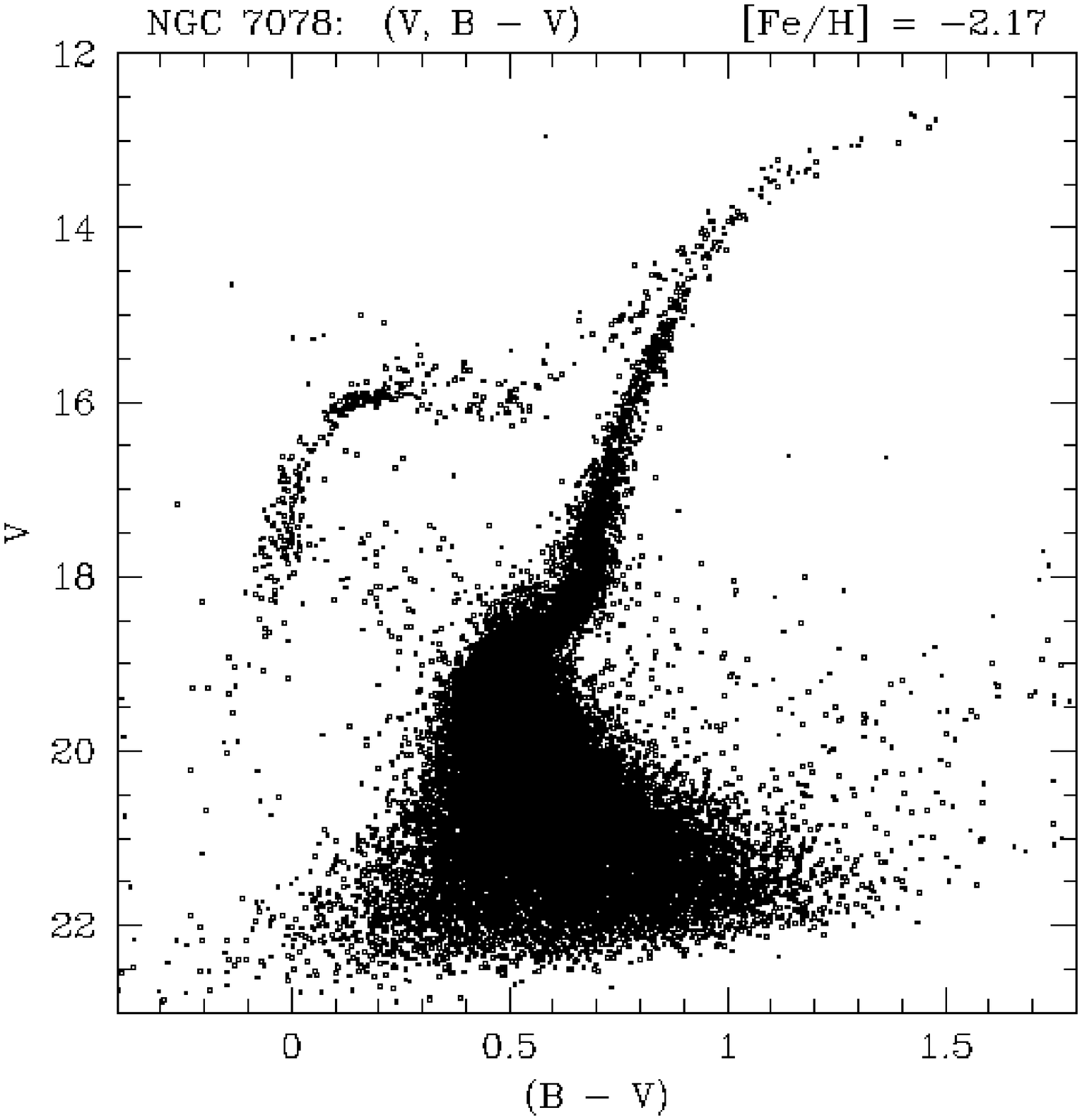,height=2.6truein,width=2.6truein}
}
\vbox{\hsize 2.5 truein
\psfig{figure=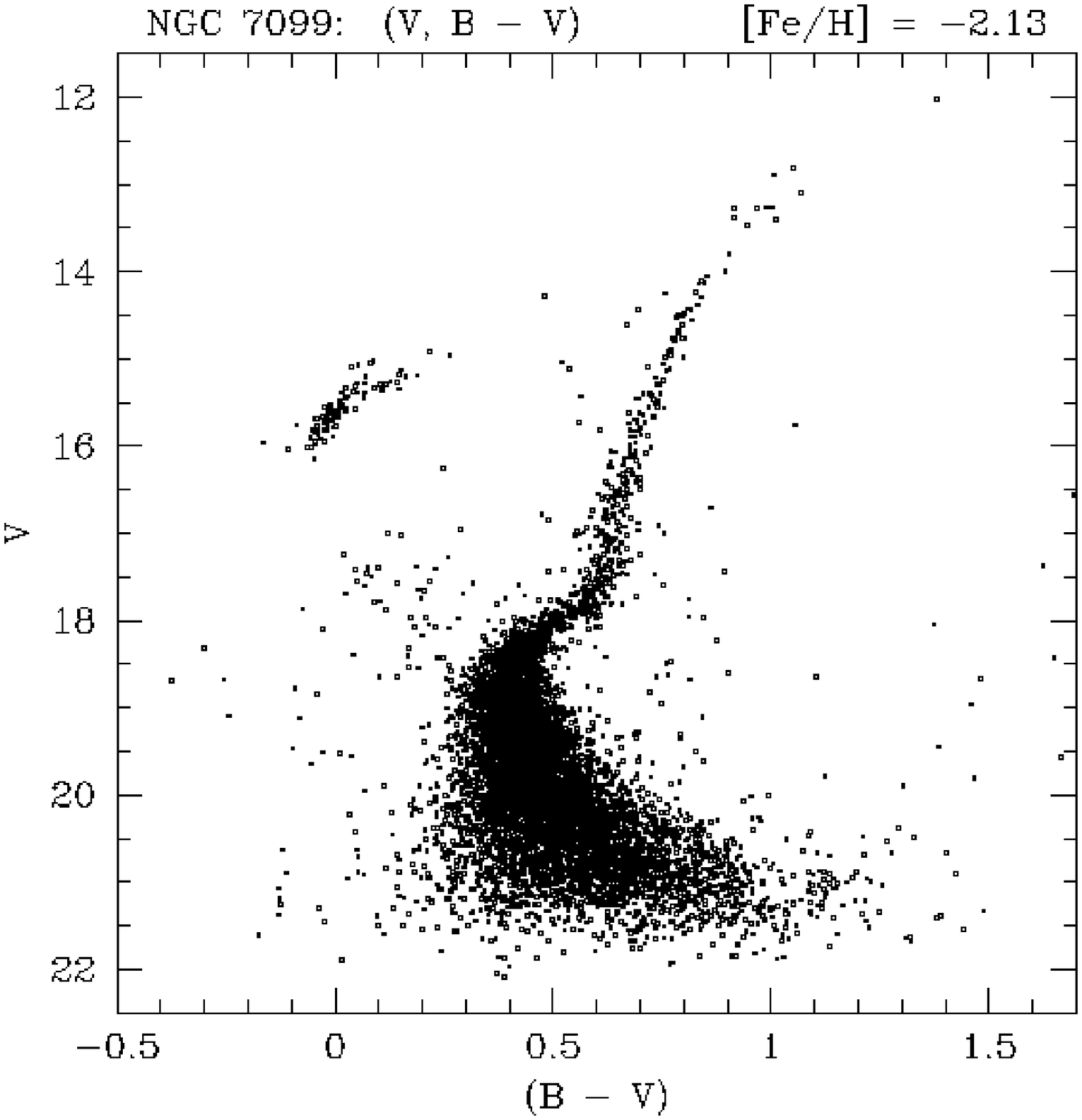,height=2.6truein,width=2.6truein}
}
}
\cptn{Figure 3. CMDs of eight more globular clusters.}

\end{document}